\documentclass[sigconf]{acmart}
\usepackage[binary-units,per-mode=symbol,per-symbol=p]{siunitx}
\usepackage[utf8]{inputenc}
\usepackage{adjustbox}
\usepackage{multirow}
\usepackage[ruled,vlined]{algorithm2e}
\usepackage{algorithmic}
\usepackage{graphicx}
\usepackage{textcomp}
\usepackage{xcolor}
\usepackage{balance}
\usepackage{enumitem}
\usepackage{setspace}
\usepackage{wrapfig}
\usepackage{listings}
\usepackage{color}
\usepackage{caption}
\usepackage{subcaption}
\usepackage{makecell}

\usepackage[table]{xcolor}
\SetKwInput{KwInput}{Input}
\SetKwInput{KwOutput}{Output}

\usepackage{amsmath}

\usepackage{mathtools}

\newcommand{\etal}{\emph{et~al.}\xspace}

\newcommand{\eg}{\emph{e.g.}, }

\newcommand{\HEVC}{\emph{High Efficiency Video Coding }}

\newcommand{\VVC}{\emph{Versatile Video Coding }}

\newcommand{\PRA}{\texttt{PRA} }
\newcommand{\CRC}{\texttt{CRC} }

\copyrightyear{2026}
\acmYear{2026}
\setcopyright{cc}
\setcctype{by}
\acmConference[MHV '26]{Mile-High Video Conference}{February 02--05, 2026}{Denver, CO, USA}
\acmBooktitle{Mile-High Video Conference (MHV '26), February 02--05, 2026, Denver, CO, USA}
\acmPrice{}
\acmDOI{10.1145/3789239.3793270}
\acmISBN{979-8-4007-2466-4/2026/02}

\begin{document}

\title{Fast Multirate Encoding for 360° Video in OMAF Streaming Workflows}

\author{Amritha Premkumar, Christian Herglotz}
\affiliation{
  \institution{Chair of Computer Engineering}
  \institution{\small{Brandenburg University of Technology Cottbus-Senftenberg}}
  \city{Cottbus}
  \country{Germany}
}

\renewcommand{\shortauthors}{Amritha Premkumar~\etal}

\begin{abstract}
Preparing high-quality $360^{\circ}$ video for HTTP Adaptive Streaming requires encoding each sequence into multiple representations spanning different resolutions and quantization parameters (QPs). For ultra-high-resolution immersive content such as 8K $360^{\circ}$ video, this process is computationally intensive due to the large number of representations and the high complexity of modern codecs. This paper investigates fast multirate encoding strategies that reduce encoding time by reusing encoder analysis information across QPs and resolutions. We evaluate two cross-resolution information--reuse pipelines that differ in how reference encodes propagate across resolutions: (i) a strict HD~$\rightarrow$~4K~$\rightarrow$~8K cascade with scaled analysis reuse, and (ii) a resolution-anchored scheme that initializes each resolution with its own highest-bitrate reference before guiding dependent encodes. In addition to evaluating these pipelines on standard equirectangular projection content, we also apply the same two pipelines to cubemap-projection (CMP) tiling, where each $360^{\circ}$ frame is partitioned into independently encoded tiles. CMP introduces substantial parallelism, while still benefiting from the proposed multirate analysis-reuse strategies. Experimental results using the SJTU 8K $360^{\circ}$ dataset show that hierarchical analysis reuse significantly accelerates HEVC encoding with minimal rate--distortion impact across both equirectangular and CMP-tiled content, yielding encoding-time reductions of roughly $33\%$--$59\%$ for ERP and about $51\%$ on average for CMP, with Bjøntegaard Delta Encoding Time (BDET) gains approaching $-50\%$ and wall-clock speedups of up to $4.2\times$.
\end{abstract}

\ccsdesc[500]{Information systems~Multimedia streaming}

\keywords{360-degree Video, Fast Multirate Encoding, HEVC, Cubemap Projection, Adaptive Streaming}

\maketitle

\section{Introduction}
Immersive media is central to virtual reality (VR), augmented reality (AR), telepresence, education, and interactive entertainment. A common format is $360^{\circ}$ video, which captures the full spherical environment and enables users to explore scenes with three rotational degrees of freedom~\cite{yaqoob_survey_2020,zink_scalable_2019}. Delivering such content at high perceptual quality is challenging, as it demands ultra-high resolutions (4K–8K and beyond), high frame rates (60–120\,fps), and specialized projection formats, imposing substantial computational and bandwidth burdens across capture, projection, compression, and streaming~\cite{huang_360-degree_2014,he_content-adaptive_2018,carreira_versatile_2020,chen_energy-efficient_2023}.

Modern immersive streaming relies on HTTP Adaptive Streaming (HAS) methods such as MPEG-DASH and HLS~\cite{mpeg_dash_ref,HLS}, where each video is encoded into a bitrate ladder comprising multiple representations at different resolutions, bitrates, and QPs~\cite{bentaleb_survey_2019,menon_ref,menon_jnd-aware_2024}. Clients switch among these representations to maintain uninterrupted playback. For $360^{\circ}$ content, adaptivity is often paired with viewport-dependent or tile-based delivery so only the user’s field of view is streamed at high quality~\cite{monakhov_data_2019,shinohara_performance_2019,yaqoob_combined_2021,qian_fbra360_2023,tang_online_2022}. Prior work spans viewport prediction, saliency, bitrate allocation, energy efficiency, and reinforcement learning–based adaptation~\cite{feng_adaptive_2025,yang_360spred_2024,khan_transusers_2024,orduna_content-immersive_2023,tu_pstile_2023,chen_energy-efficient_2023,jiang_reinforcement_2021,qian_fbra360_2023, katsenou2026multiobjectiveparetofrontoptimizationefficient}, but most approaches assume that full multirate representation sets are already available, overlooking the substantial cost of preparing the underlying multirate representations—a gap our work explicitly targets~\cite{10.1145/3638036.3640801} and which prior $360^{\circ}$ HAS studies do not address.

\begin{figure}[t]
\centering
\includegraphics[width=0.485\columnwidth]{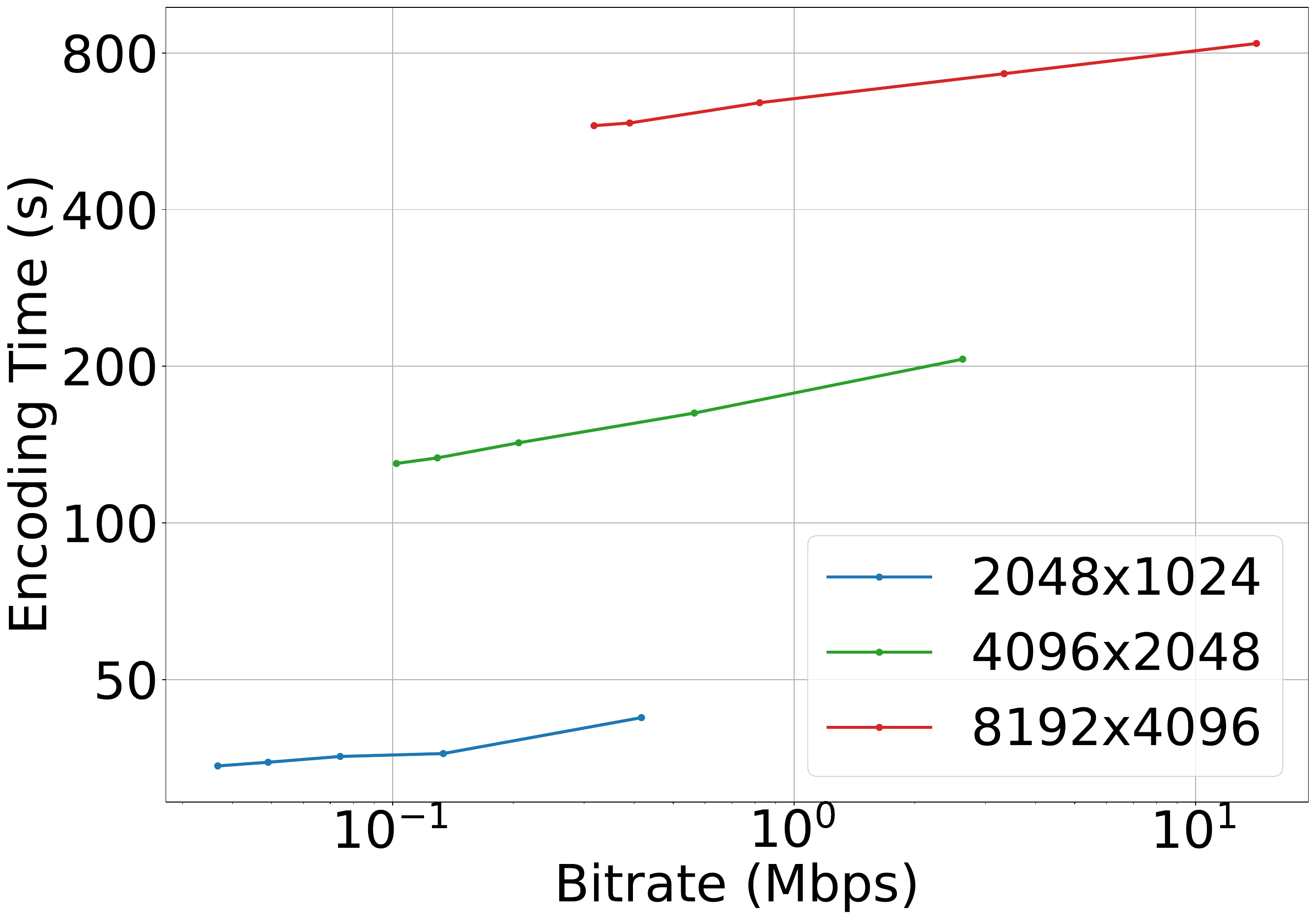}
\includegraphics[width=0.485\columnwidth]{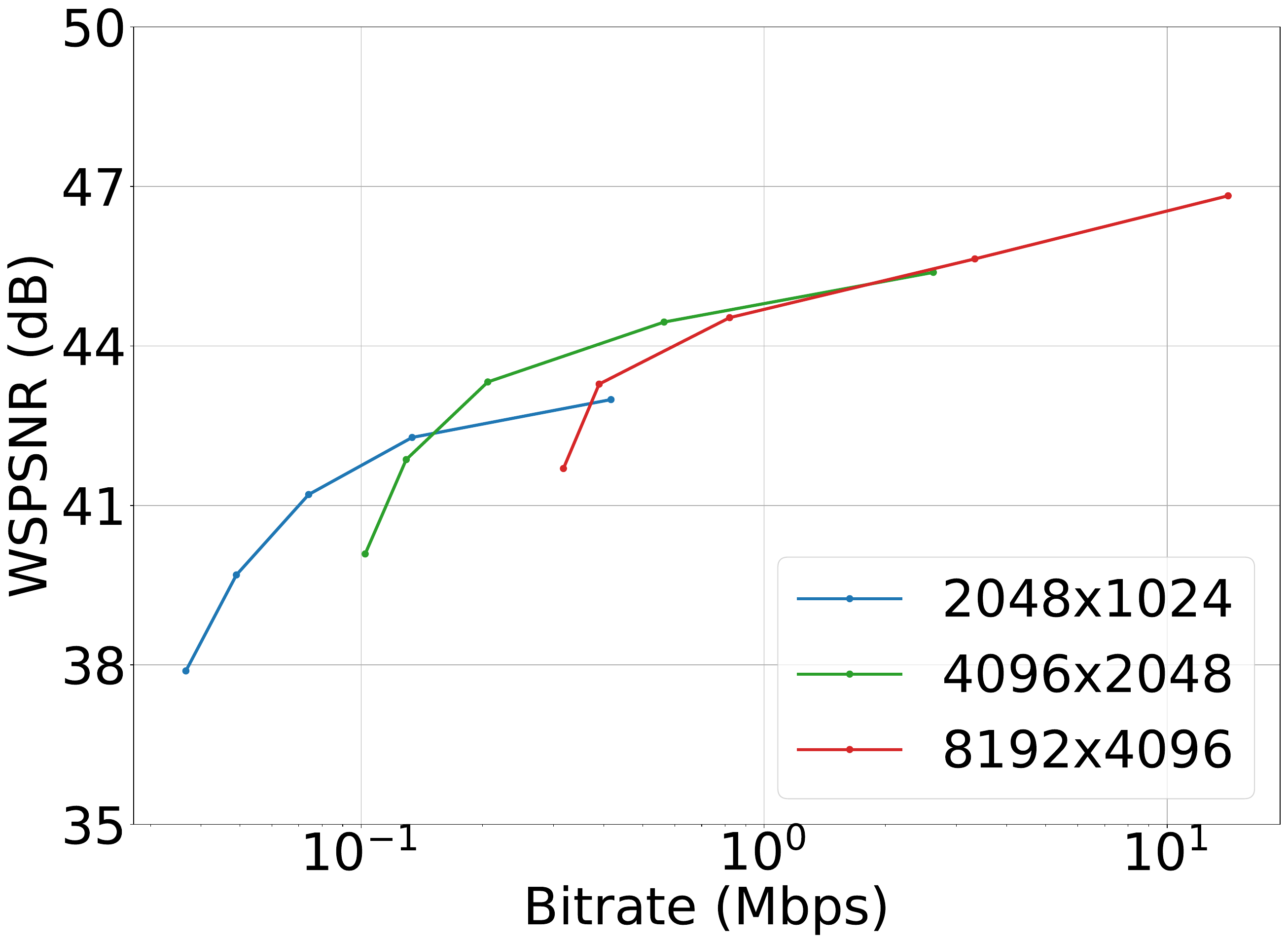}
\caption{Rate-quality, and rate-encoding time curves of \emph{Kaixuanmenye pano}~\cite{liu_sjtu_2017}. Here, quality is represented using WSPSNR.}
\vspace{-2.98em}
\label{fig:intro_rd_res}
\end{figure}

This challenge is acute in OMAF workflows~\cite{Hannuksela_ref}, which define projection metadata, viewport-adaptive signaling, and region-based mechanisms such as tiling or cubemap decomposition~\cite{cortes_influence_2020,monakhov_data_2019,yang_enhancing_2021}. These features improve delivery efficiency but require many encoded bitstreams across resolutions and regions. Encoding 15 representations of an 8K $360^{\circ}$ video can exceed 30 CPU-hours on commodity hardware~\cite{kufa_software_2023}, making exhaustive full-search encoding impractical for large VR libraries or latency-sensitive scenarios such as live streaming~\cite{singh_aerial_2023}. Recent work has further highlighted the importance of encoding latency and energy consumption as first-class optimization objectives in adaptive video streaming workflows \cite{10.1145/3638036.3640801}. Figure~\ref{fig:intro_rd_res} illustrates this computational cost: encoding times rise sharply with resolution, while WSPSNR gains saturate unless bitrate increases significantly. Preparing a full ladder therefore requires repeated, expensive encodes, motivating fast multirate encoding techniques.

Fast multirate and multiresolution encoding exploits a key observation: encoder decisions are highly correlated across representations of the same content. Prior work~\cite{carreira_versatile_2020,wang_fast_2025,he_content-adaptive_2018} shows that CTU partitioning, intra/inter mode choices, motion estimation, and transform decisions remain structurally stable across QPs and resolutions. Reusing analysis from previously encoded representations can therefore avoid many redundant searches, substantially reducing encoding time with modest RD impact~\cite{Schroeder_ref,menon2025blockpartitioning_vvc}. While x265 exposes analysis‑reuse primitives~\cite{menon_ref,aruna_ref}, their behavior for $360^{\circ}$ layouts, OMAF‑driven bitrate ladders, and cross‑resolution pipelines has not been characterized; our work provides the first projection‑aware evaluation of reuse flows.

\begin{figure}[t]
\centering
\begin{subfigure}{0.47\columnwidth}
\centering
\includegraphics[width=\textwidth]{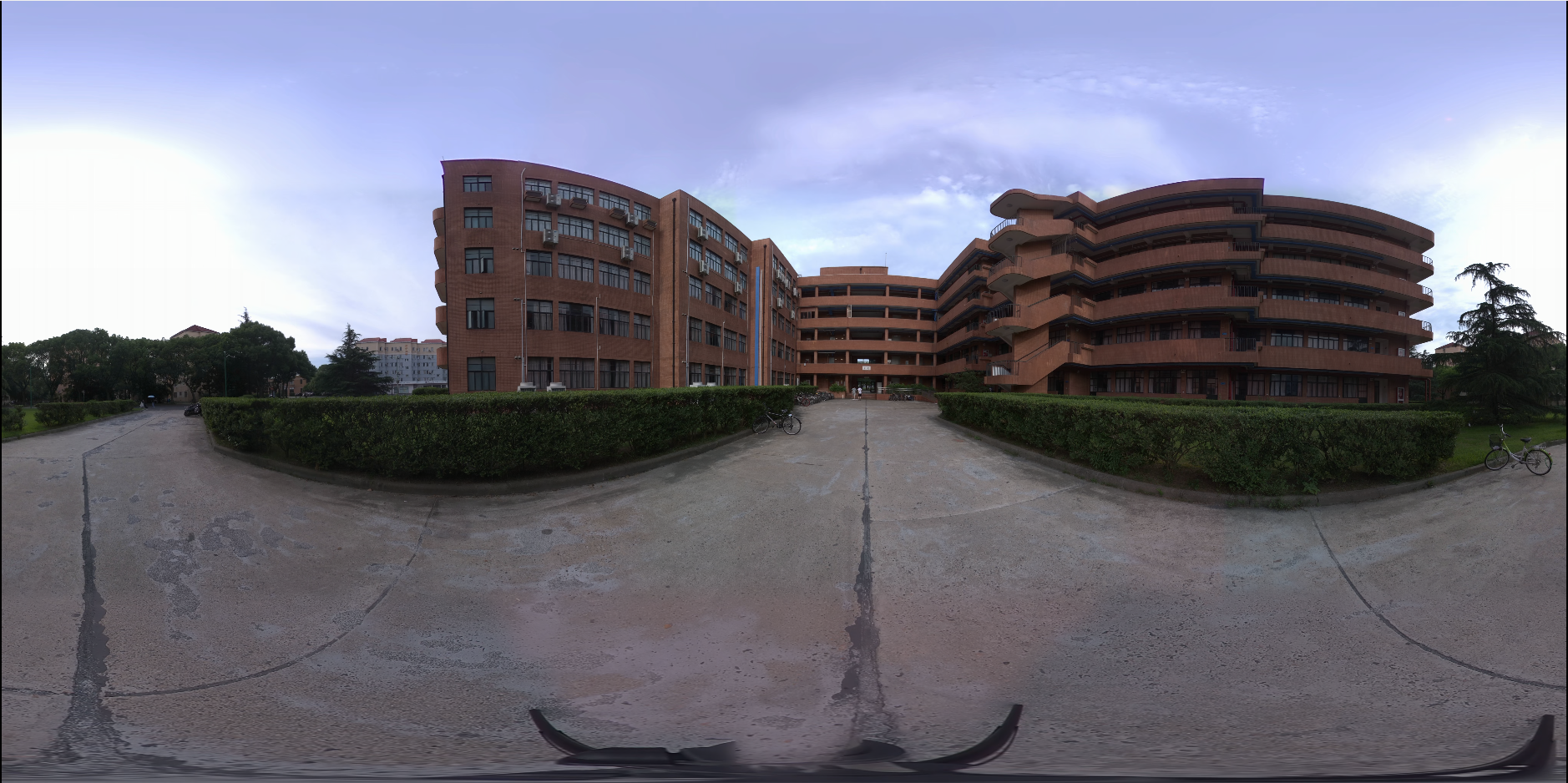}
\caption{ERP}
\end{subfigure}
\begin{subfigure}{0.47\columnwidth}
\centering
\includegraphics[width=\textwidth]{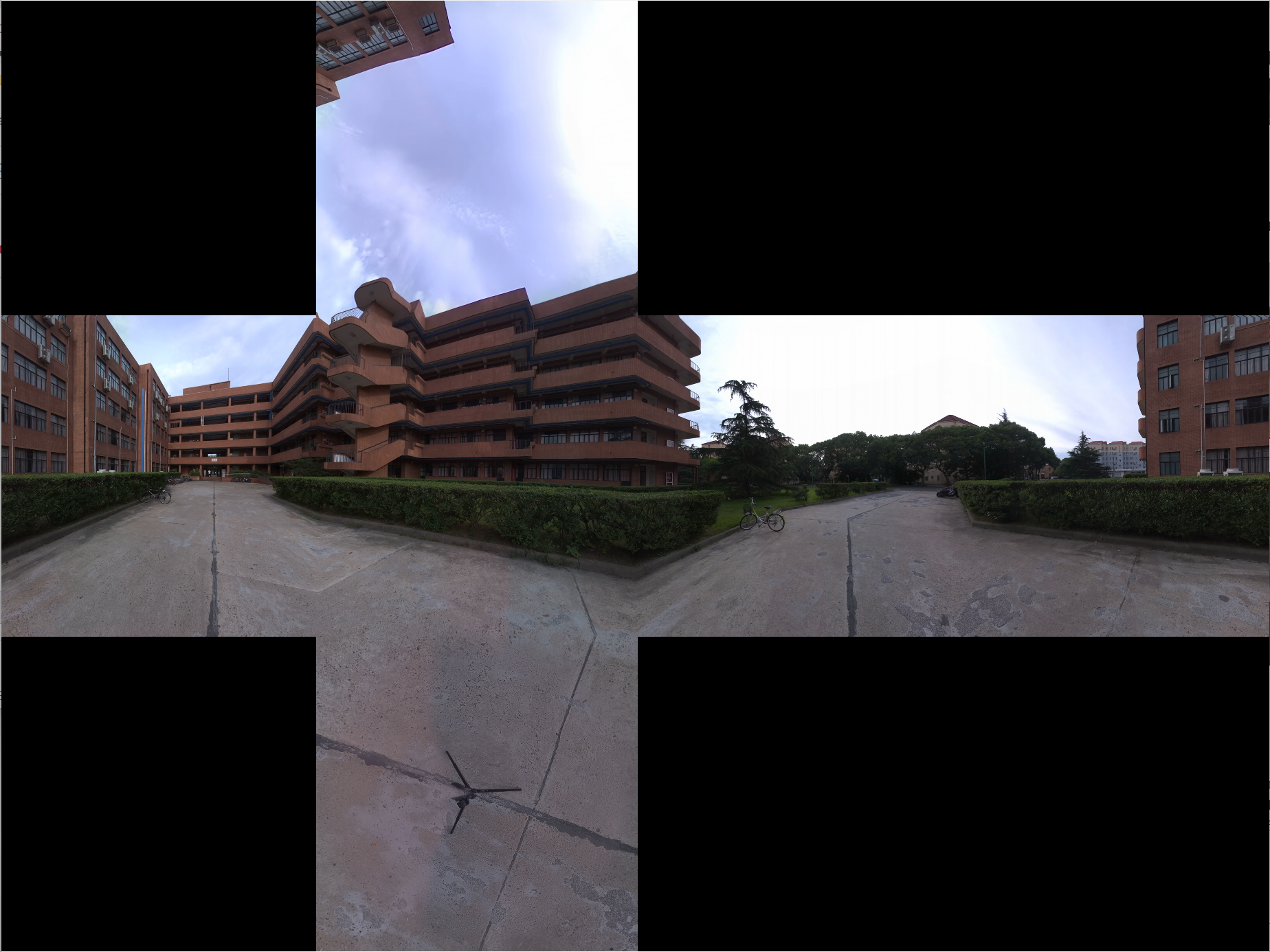}
\caption{CMP}
\end{subfigure}
\vspace{-0.5em}
\caption{Comparison of 360° Projection Formats: ERP versus CMP.}
\vspace{-1.5em}
\label{fig:erp_cmp}
\end{figure}

Applying such reuse to spherical video, however, is challenging. Equirectangular projection (ERP), the dominant $360^{\circ}$ layout, introduces strong geometric distortions near the poles~\cite{lowcomp_erp_encoder,he_geometry_2017}, affecting spatial and motion statistics and weakening cross-resolution consistency. Rotational motion, parallax, and viewport dynamics further increase temporal variability~\cite{nguyen_impact_2017,cortes_influence_2020}. Cubemap projection (CMP) offers a more uniform alternative: it decomposes the sphere into six square faces (Fig.~\ref{fig:erp_cmp}), reducing geometric distortion and yielding more homogeneous block statistics~\cite{sreedhar_viewport-adaptive_2016,shinohara_performance_2019}. CMP also aligns naturally with OMAF's region-based design, enabling independent, parallel face-wise encoding and amplifying the benefits of analysis reuse.

This work makes the following key contributions:
\begin{enumerate}
\item \emph{Projection-Aware Multirate Encoding for 360° Video:} We extend multirate and multiresolution analysis‑reuse methods to spherical video, addressing ERP distortions and viewport-driven variability, and show that CMP improves cross-resolution reuse stability.
\item \emph{Comprehensive Evaluation on 8K 360° Content:} Using the SJTU 8K $360^{\circ}$ dataset~\cite{liu_sjtu_2017}, we quantify encoding-time reductions, RD trade-offs, and scalability of the proposed techniques in OMAF-compliant pipelines, demonstrating substantial speedups with minimal RD loss.
\end{enumerate}

\emph{\textbf{Paper Outline:}} Section~\ref{sec:background} reviews related work. Section~\ref{sec:method} describes the proposed architecture, Section~\ref{sec:exp_setup} details the experimental setup, Section~\ref{sec:results} presents the results, and Section~\ref{sec:conc} concludes with future directions.

\section{Fast Multirate Encoding}
\label{sec:background}
Generating full bitrate ladders for HAS is computationally expensive, especially for modern codecs like \HEVC~(HEVC) and \VVC~(VVC)~\cite{HEVC, vvc_ref, hevc_vvc_enc_comp}. Fast multirate encoding addresses this by reusing encoder decisions-- CTU partitioning, prediction modes, motion vectors, and transform choices—across representations of the same video, exploiting their strong correlation across QPs and resolutions~\cite{Schroeder_ref,Schroeder_multires,menon_ref}. This reuse can cut encoding time dramatically with minimal rate–distortion (RD) loss. Two main paradigms exist: \emph{top-down} and \emph{bottom-up}.

\subsection{Top-Down vs. Bottom-Up Paradigms}
The \emph{top-down} approach propagates decisions from the highest-quality representation (lowest QP or highest resolution) to others~\cite{Schroeder_ref, Schroeder_multires}. While RD performance is strong, serialization limits parallelism, making it unsuitable for time-sensitive or large-scale workflows. Conversely, the \emph{bottom-up} approach starts from the fastest-to-encode representation (lowest resolution or highest QP)~\cite{Goswami_ref,menon_ref,menon2025blockpartitioning_vvc}. Analysis becomes available early, enabling high concurrency and scalability. Although RD loss is slightly higher than top-down, bottom-up approaches are far more practical for VR/360° pipelines requiring dozens of representations.

\subsection{Cross-Resolution and Cross-QP Prediction}
Encoder analysis transfers well across neighboring QPs and resolutions. Cross-QP reuse exploits stable block-partitioning patterns, while cross-resolution reuse leverages spatial self-similarity~\cite{menon2025blockpartitioning_vvc,Schroeder_multires}. Mapping decisions across resolutions, however, requires precise scaling of CTU coordinates and block sizes. Recent heuristics and lightweight ML models improve robustness, but ultra-high-resolution encoding (4K–8K) still incurs significant complexity, motivating further optimization.

\begin{figure*}[t]
\centering
\includegraphics[width=0.695\textwidth]{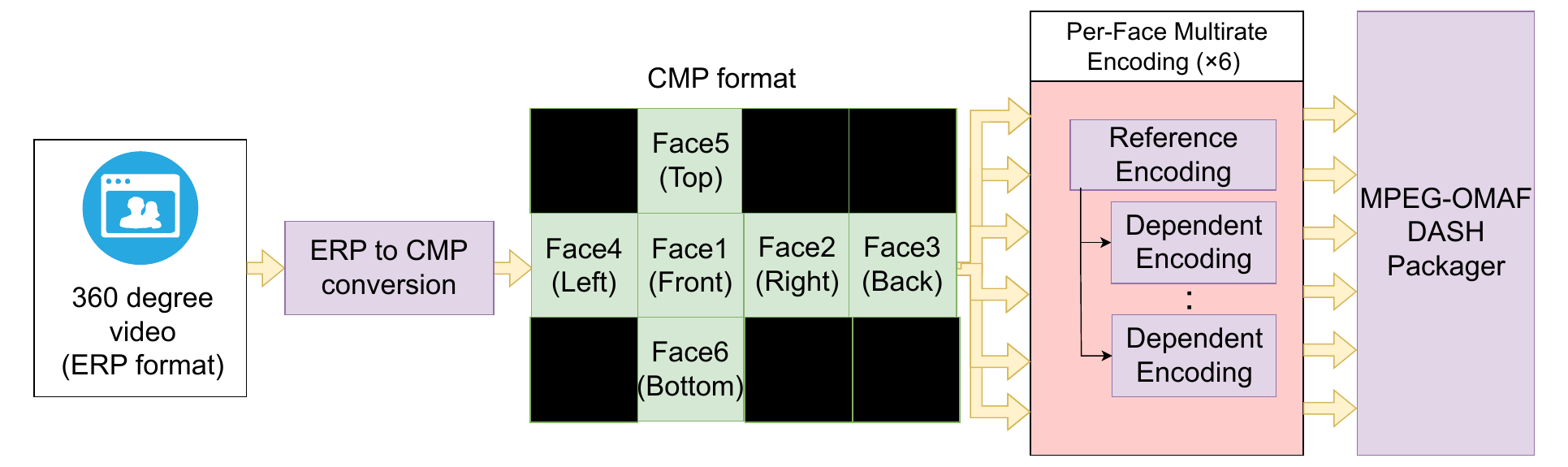}
\caption{Overview of the proposed fast multirate encoding pipeline. The optional ERP$\rightarrow$CMP conversion is applied only in CMP-\CRC and CMP-\PRA. For ERP-\CRC and ERP-\PRA, the pipeline omits the projection box; the multirate reuse logic (anchor + dependents, analysis save/load) remains identical.}
\vspace{-0.5em}
\label{fig:method_arch}
\end{figure*}

\subsection{Challenges for 360$^{\circ}$ Video}
Spherical video complicates reuse due to nonuniform pixel density. ERP, the dominant layout, introduces severe polar distortion, altering spatial and motion statistics and reducing transfer reliability. Consequently, existing multirate methods degrade when applied to 360° content. CMP mitigates these issues by decomposing the sphere into six square faces with uniform characteristics. CMP improves motion estimation, stabilizes block statistics, and enables independent face-wise encoding, unlocking massive parallelism. For OMAF-style pipelines requiring dozens of representations, these properties make CMP essential for practical, fast multirate encoding.

Without projection-aware acceleration, preparing full bitrate ladders for 8K 360° video remains prohibitively slow, motivating software‑based fast multirate pipelines; hardware encoders offer high throughput but generally lack CTU‑level analysis export/import, and therefore cannot participate in cross‑QP or cross‑resolution reuse. Our method addresses this gap by combining CMP with efficient cross-resolution and cross-QP reuse, delivering scalability without sacrificing RD performance.

\section{Proposed Framework}
\label{sec:method}
Our framework takes an input $360^\circ$ ERP video and either (i) encodes it directly, or (ii) converts it to a CMP layout to reduce projection distortion and expose additional parallelism. We then perform fast multirate encoding by fully encoding a chosen \emph{anchor} and reusing its analysis to accelerate the remaining \emph{dependent} encodes. All resulting bitstreams are finally packaged into an OMAF-compliant DASH presentation (Figure~\ref{fig:method_arch}).

We evaluate four practical variants by combining the projection format (ERP or CMP) with the analysis-sharing strategy (CRC or \PRA):
\begin{itemize}
    \item \textbf{ERP-\CRC}: ERP projection with a cascaded HD$\rightarrow$4K$\rightarrow$8K flow.
    \item \textbf{ERP-\PRA}: ERP projection with per-resolution anchors.
    \item \textbf{CMP-\CRC}: CMP projection with a cascaded flow applied per face.
    \item \textbf{CMP-\PRA}: CMP projection with per-face, per-resolution anchors.
\end{itemize}
ERP variants operate on a single full-frame tile, while CMP variants reuse analysis independently per face, unlocking more parallelism and benefiting from CMP’s more uniform sampling.

At each resolution layer (HD, 4K, 8K), we select one \emph{anchor} representation—low (LQ), medium (MQ), or high (HQ) quality—to run full RDO. Its decisions are reused to accelerate all \emph{dependent} representations at the same resolution and, in CRC, at higher resolutions.

\subsection{Spherical-to-Cubemap Projection}

\paragraph*{Formulation}
Let the $360^\circ$ content be defined on the sphere $\mathbb{S}^2$ by a radiance function
\begin{equation}
\mathcal{V} : \mathbb{S}^2 \times \mathbb{N} \rightarrow \mathbb{R}^3,
\qquad
\mathcal{V}(\boldsymbol{\omega}, t) = (Y, Cb, Cr),
\end{equation}
where $\boldsymbol{\omega}$ is a viewing direction and $(Y,Cb,Cr)$ are YCbCr components. The ERP projection samples this function via
\begin{equation}
E(u,v,t) = \mathcal{V}(\boldsymbol{\omega}_{\mathrm{ERP}}(u,v), t), \qquad (u,v)\in[0,W)\times[0,H).
\end{equation}

The CMP layout maps $\mathbb{S}^2$ to six square domains $\Omega_k$ via
\begin{equation}
F_k(x,y,t) = \mathcal{V}(\boldsymbol{\omega}_k(x,y), t), \qquad (x,y)\in\Omega_k,
\end{equation}
with $\boldsymbol{\omega}_k(\cdot)$ denoting the face-specific inverse projection. CMP reduces geometric distortion and stabilizes encoder statistics, improving the reliability of cross-resolution reuse.

\subsection{Rate Ladder and Representation Structure}
\paragraph*{Intuition} We encode a compact set of quality levels (CRFs) per face. Only the anchor performs full RDO; dependents reuse its analysis to skip most partitioning, mode, and motion searches.

\paragraph*{Formulation} For each face $k$, the CRF ladder is
\begin{equation}
\mathcal{Q}=\{q_0,q_1,\dots,q_{L-1}\}, \qquad q_0<q_1<\dots<q_{L-1},
\end{equation}
with $q_0$ denoting the highest-quality anchor. Representation $\ell$ is encoded as
\begin{equation}
R_{k,\ell} = \mathcal{E}(F_k; q_\ell), \qquad \ell\in\{0,\dots,L-1\},
\end{equation}
where $\mathcal{E}(\cdot)$ denotes x265 encoding with analysis saving/loading. Dependents load and reuse the anchor’s analysis to accelerate encoding.

\subsection{Reference Encoding: Full RDO and Analysis Extraction}
The anchor at each resolution (or face) performs full RDO, exhaustively searching block partitions, prediction modes, and motion vectors. Optimal decisions are stored per frame and CU:
\begin{equation}
d_{k,0}^{(n,i)}=\big(s_{k,0}^{(n,i)},\,m_{k,0}^{(n,i)},\,\mathbf{v}_{k,0}^{(n,i)}\big),
\end{equation}
where $s$ is the split pattern, $m$ the prediction mode, and $\mathbf{v}$ the motion vector.

\subsection{Dependent Encodings via Analysis Reuse}
\paragraph*{Overview}
Dependent encodes reuse the anchor’s decisions as strong priors, reducing the search space and avoiding expensive evaluations.

\paragraph*{Formulation}
For each $\ell\ge1$:
\begin{equation}
\mathbf{d}_{k,\ell} =
\arg\min_{\mathbf{d}\in\mathcal{S}_{k,0}}
\big( D(\mathbf{d}) + \lambda(q_\ell)\,R(\mathbf{d}) \big),
\end{equation}
where admissible decisions satisfy:
\begin{itemize}
    \item \textbf{Split reuse:} $s_{k,\ell}^{(n,i)} \preceq s_{k,0}^{(n,i)}$ (no deeper splits).
    \item \textbf{Mode reuse:} $m_{k,\ell}^{(n,i)} = m_{k,0}^{(n,i)}$ (skip mode search).
    \item \textbf{MV reuse:} $\mathbf{v}_{k,\ell}^{(n,i)}\!=\!\mathbf{v}_{k,0}^{(n,i)}$ (perform only local refinement).
\end{itemize}
For cross-resolution reuse (HD$\rightarrow$4K, 4K$\rightarrow$8K), x265 scales the anchor’s analysis to the target CU grid.

\subsection{Analysis Sharing Schemes: \CRC and \PRA}
We compare two practical multiresolution strategies:
\begin{itemize}
    \item \textbf{CRC:} A strict bottom-up cascade (HD$\rightarrow$4K$\rightarrow$8K). Each resolution has a single anchor whose analysis accelerates all dependents and seeds the next resolution. This maximizes cross-resolution reuse at minimal memory cost.
    \item \textbf{PRA:} Each resolution independently selects an anchor (LQ/MQ/HQ). Dependents reuse this anchor, and anchors assist in initializing the next resolution. PRA offers greater flexibility but stores one analysis file per resolution.
\end{itemize}

\subsection{OMAF Packaging and Output}
Encoded ERP or CMP representations are packaged into an OMAF-compliant DASH presentation, including projection metadata, region signalling, and representation descriptors, enabling seamless integration into standard viewport-adaptive players.

\begin{table}[t]
\centering
\caption{Framework variants combining projection and analysis-sharing strategy.}
\resizebox{0.9915\linewidth}{!}{
\begin{tabular}{l| l| l}
\specialrule{.12em}{.05em}{.05em}
\specialrule{.12em}{.05em}{.05em}
\textbf{Variant} & \textbf{Projection} & \textbf{Strategy (flow)} \\
\specialrule{.12em}{.05em}{.05em}
\specialrule{.12em}{.05em}{.05em}
ERP-\CRC & ERP (single tile) & Cascaded HD$\rightarrow$4K$\rightarrow$8K \\
ERP-\PRA & ERP (single tile) & Per-resolution anchors (HD, 4K, 8K) \\
CMP-\CRC & CMP (6 faces)     & Cascaded HD$\rightarrow$4K$\rightarrow$8K per face \\
CMP-\PRA & CMP (6 faces)     & Per-resolution anchors per face \\
\specialrule{.12em}{.05em}{.05em}
\specialrule{.12em}{.05em}{.05em}
\end{tabular}}
\vspace{-0.95em}
\label{tab:variants}
\end{table}

ERP suffers from distortion that weakens analysis reuse, and naïve full-search encoding of all ladder levels is prohibitively slow for 8K $360^\circ$ content. By combining CMP’s uniformity and face-wise parallelism with \CRC/\PRA analysis sharing and per-resolution anchors (LQ/MQ/HQ), the framework achieves scalable multirate encoding with minimal RD loss and OMAF-ready outputs.

\section{Experimental Setup}
\label{sec:exp_setup}

\subsection{Implementation Configuration in x265}
\label{sec:x265_config}
Our framework builds on x265’s analysis–reuse subsystem~\cite{Goswami_ref,aruna_ref}. During each anchor encode, per-frame analysis is stored using \textit{analysis-save}, and is later imported by dependent encodes via \textit{analysis-load}. Reuse strength is controlled through \textit{analysis-save-reuse-level} and \textit{analysis-load-reuse-level}; we use the strongest setting (\emph{level 10}), which preserves CU-level structure, prediction modes, and motion information.

\paragraph{Cross-QP reuse (same resolution).}
Dependent encodes reuse CU splits, inter modes, and motion information with \textit{load-reuse-level=10}. Limited refinement is permitted using \textit{refine-intra=4}, \textit{refine-inter=2}, and \textit{refine-mv=1}, ensuring robustness without sacrificing speed.

\paragraph{Cross-resolution reuse.}
For HD$\rightarrow$4K and 4K$\rightarrow$8K transitions, x265 scales CTU-level decisions using \textit{scale-factor=2} together with level‑10 reuse, eliminating most partitioning, mode search, and motion estimation at higher resolutions.

\subsection{Dataset}
\begin{figure}[t]
\centering
\includegraphics[width=0.35\textwidth]{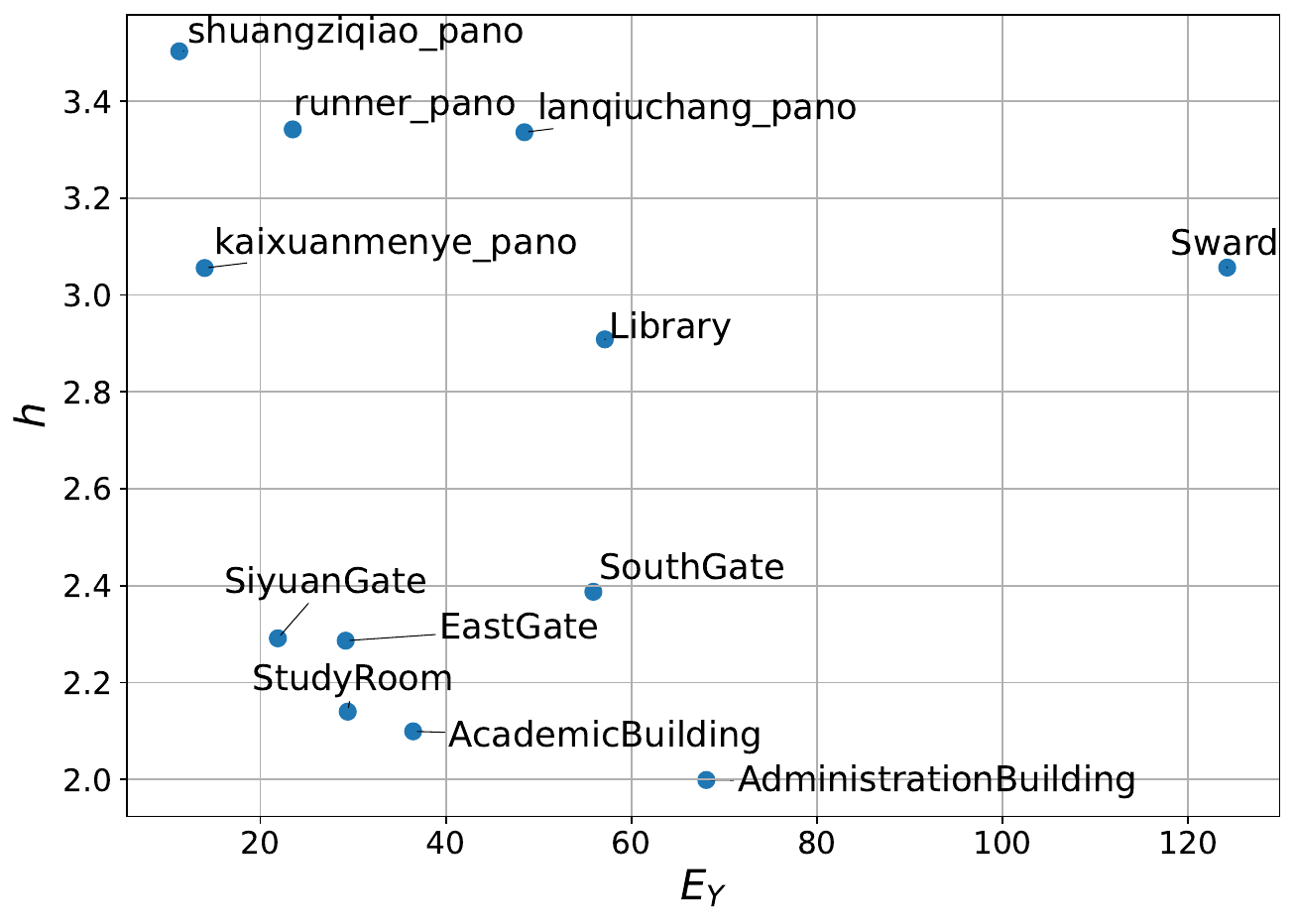}
\caption{Complexity analysis of the SJTU dataset using $(E_Y, h)$ from VCA~\cite{vca_ref}. Higher values indicate more complex motion and texture.}
\label{fig:EhL_res}
\end{figure}

We use the SJTU $360^{\circ}$ dataset~\cite{liu_sjtu_2017}, which contains 15 ERP sequences at $8192\times4096$ resolution. The clips span \SI{30}{\second} each; in our setup, a single 8K full‑search reference encode requires several minutes per CRF, underscoring the cost of preparing full bitrate ladders. The distribution of content complexity, computed via $(E_Y, h)$ features from the Video Complexity Analyzer (VCA)~\cite{vca_ref}, is shown in Figure~\ref{fig:EhL_res}.

For projection‑aware analysis and improved parallelism, each ERP sequence is converted to CMP using a standard sphere‑to‑cubemap mapping. This produces six uniformly sampled faces of size $2048\times2048$ corresponding to the \{front, back, left, right, top, bottom\}. Each face is treated as an independent tile during encoding.

All experiments are executed on a server with an Intel Xeon processor (32 threads) and 128\,GB RAM, enabling both serial runtime evaluation (per tile) and parallel runtime evaluation (six tiles concurrently). Table~\ref{tab:exp_par} summarizes resolutions, QPs, and codec settings.

\begin{table}[t]
\caption{Experimental parameters and values.}
\centering
\resizebox{0.935\columnwidth}{!}{
\begin{tabular}{l||c}
\specialrule{.12em}{.05em}{.05em}
\specialrule{.12em}{.05em}{.05em}
\emph{Parameter} & \emph{Values}\\
\specialrule{.12em}{.05em}{.05em}
\specialrule{.12em}{.05em}{.05em}
$\mathcal{R}$ & \{ 2048x1024 (HD), 4096x2048 (4K), 8192x4096 (8K) \}  \\
\hline
$\mathcal{Q}$ & \{ 22, 27, 32, 37, 42 \} \\
\hline
Target encoder & x265 [medium][intra period=\SI{1}{\second}][4 CPU threads] \\
\hline
Target decoder & HM [1 CPU thread] \\
\specialrule{.12em}{.05em}{.05em}
\specialrule{.12em}{.05em}{.05em}
\end{tabular}
}
\label{tab:exp_par}
\end{table}

\begin{figure*}[t]
\centering
\includegraphics[width=0.695\textwidth]{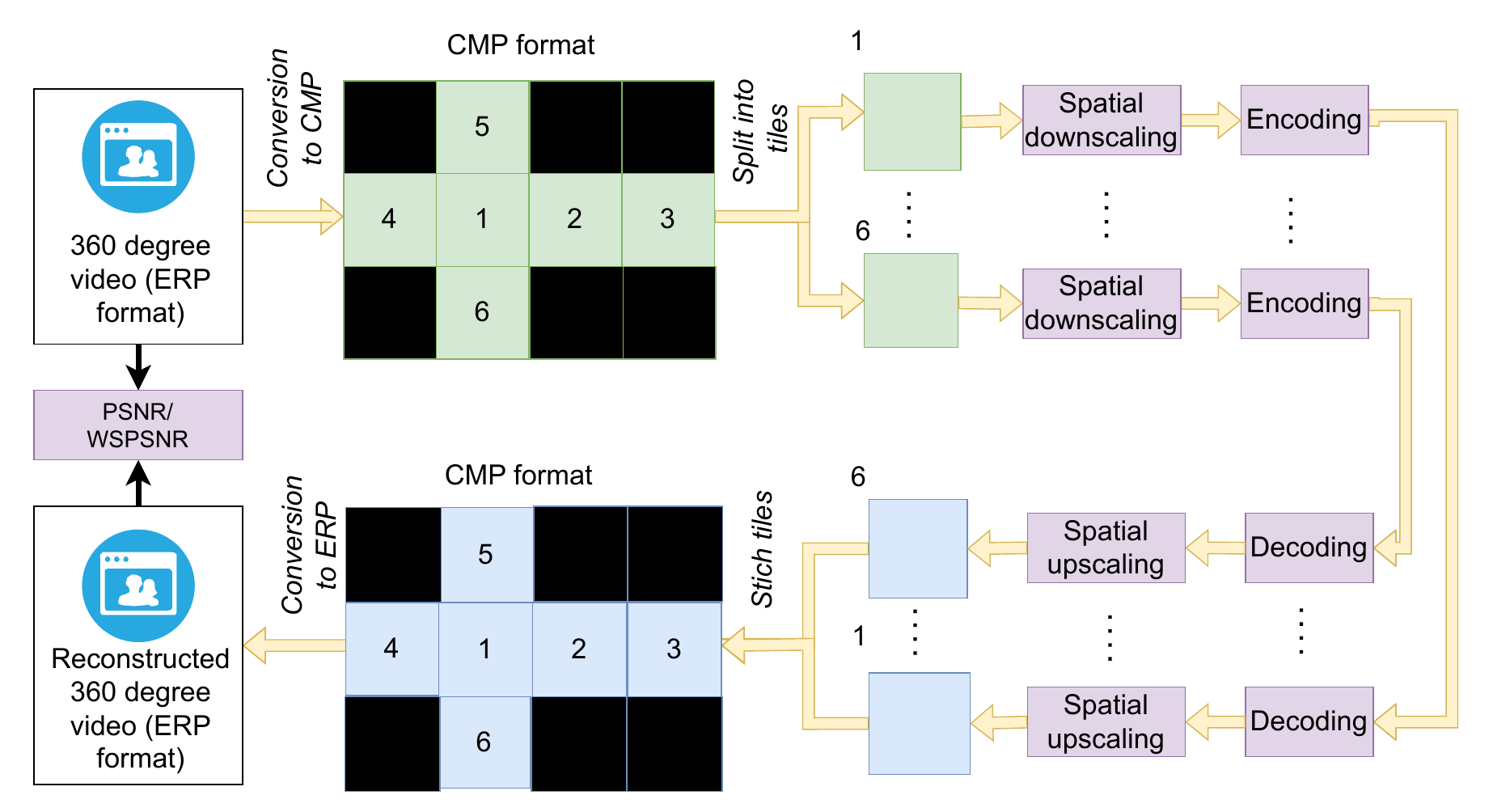}
\caption{Quality evaluation framework.}
\label{fig:q_eval}
\end{figure*}

\subsection{Metrics}
All evaluations compare each proposed method against \emph{ERP-Default}, the full-search ERP encode.

\begin{description}[font=\normalfont\itshape]
\item[Quality metrics] We compute PSNR and WSPSNR for each QP using Samsung 360tool.\footnote{\url{https://github.com/Samsung/360tools}} Figure~\ref{fig:q_eval} illustrates the evaluation pipeline.

\item[Serial encoding runtime difference ($\Delta T_{\text{S}}$)]
Assuming sequential encoding, the relative runtime change is
\begin{equation}
\Delta T_{\text{S}}=
\frac{\sum \tau_{\text{method}} - \sum \tau_{\text{ref}}}
{\sum \tau_{\text{ref}}}.
\end{equation}

\item[Parallel encoding runtime difference ($\Delta T_{\text{P}}$)]
Assuming concurrent encoding, runtime is dominated by the slowest representation:
\begin{equation}
\Delta T_{\text{P}}=
\frac{\max(\tau_{\text{method}}) - \max(\tau_{\text{ref}})}
{\max(\tau_{\text{ref}})}.
\end{equation}

\item[\textbf{Bjøntegaard Delta Metrics}]
BD-PSNR and BD-WSPSNR measure average quality differences relative to ERP‑Default~\cite{DCC_BJDelta}, where positive values denote gains and negative values small losses. BDET~\cite{Herglotz_2019} evaluates relative encoding-time savings at equal quality; negative BDET indicates faster encoding for the same RD performance.
\end{description}

\section{Results and Discussion}
\label{sec:results}
\begin{figure}[t]
\centering
\begin{subfigure}{0.49\textwidth}
\centering
\includegraphics[width=0.45\columnwidth]{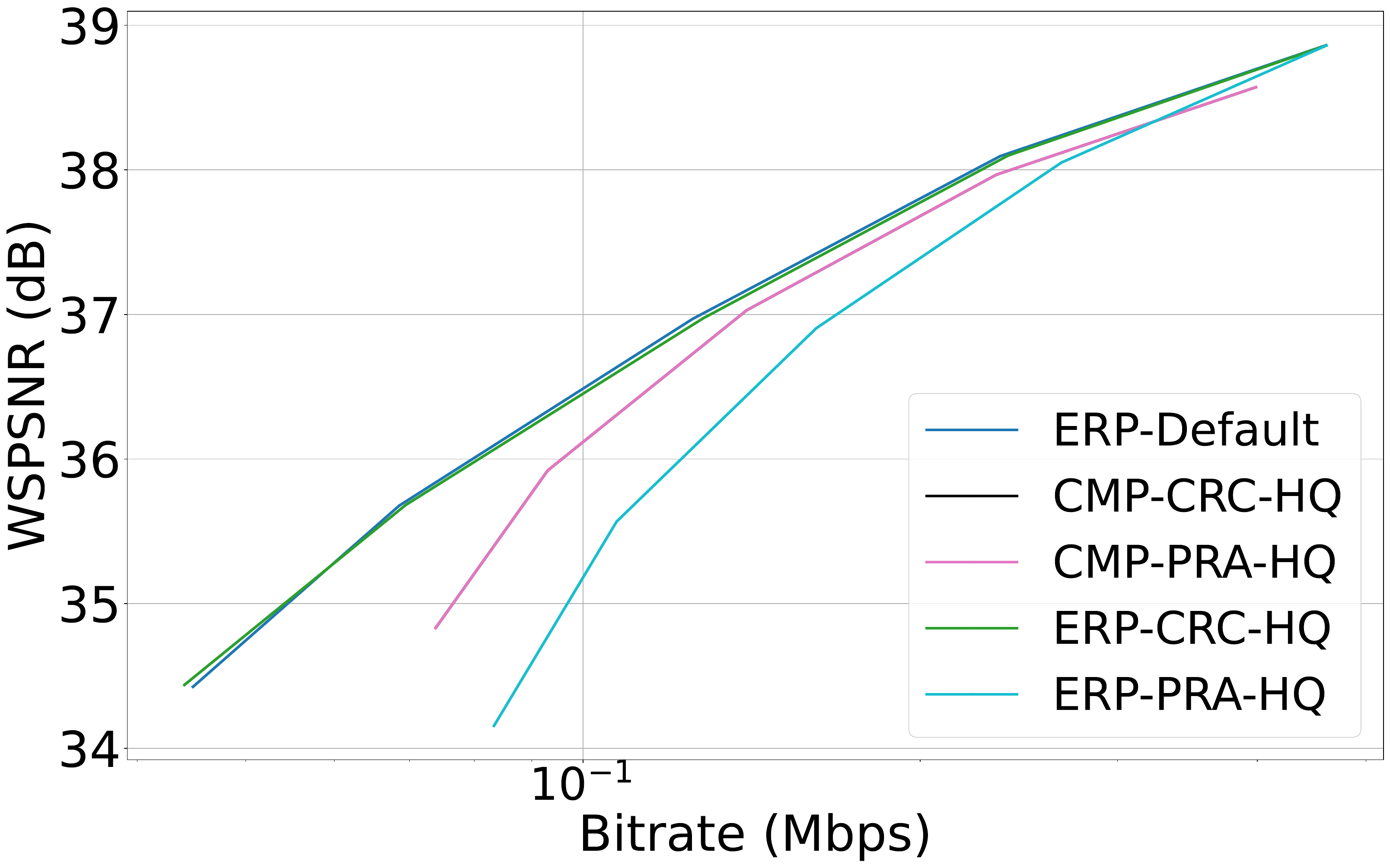}
\includegraphics[width=0.45\columnwidth]{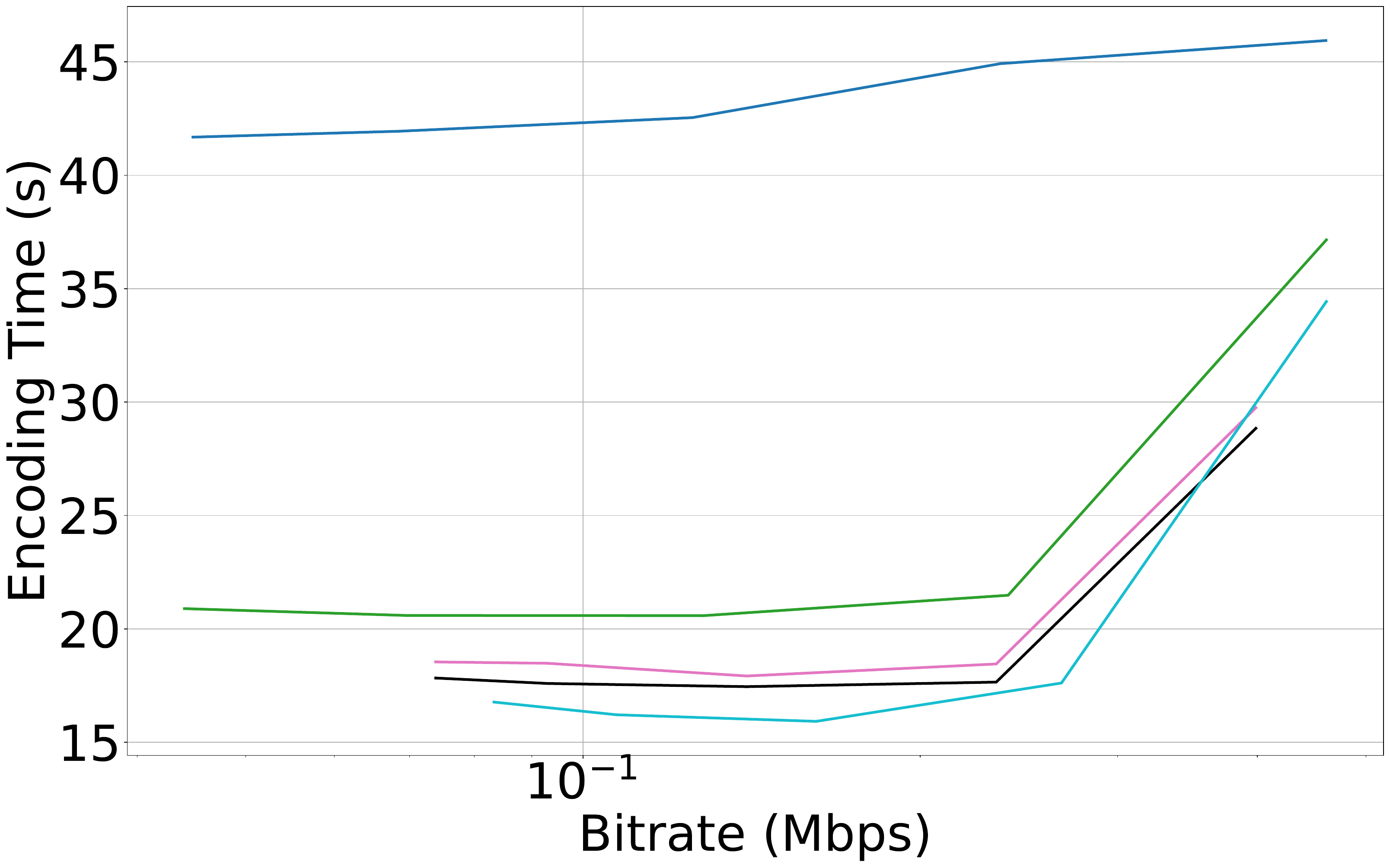}
\caption{HD}
\end{subfigure}

\begin{subfigure}{0.49\textwidth}
\centering
\includegraphics[width=0.45\columnwidth]{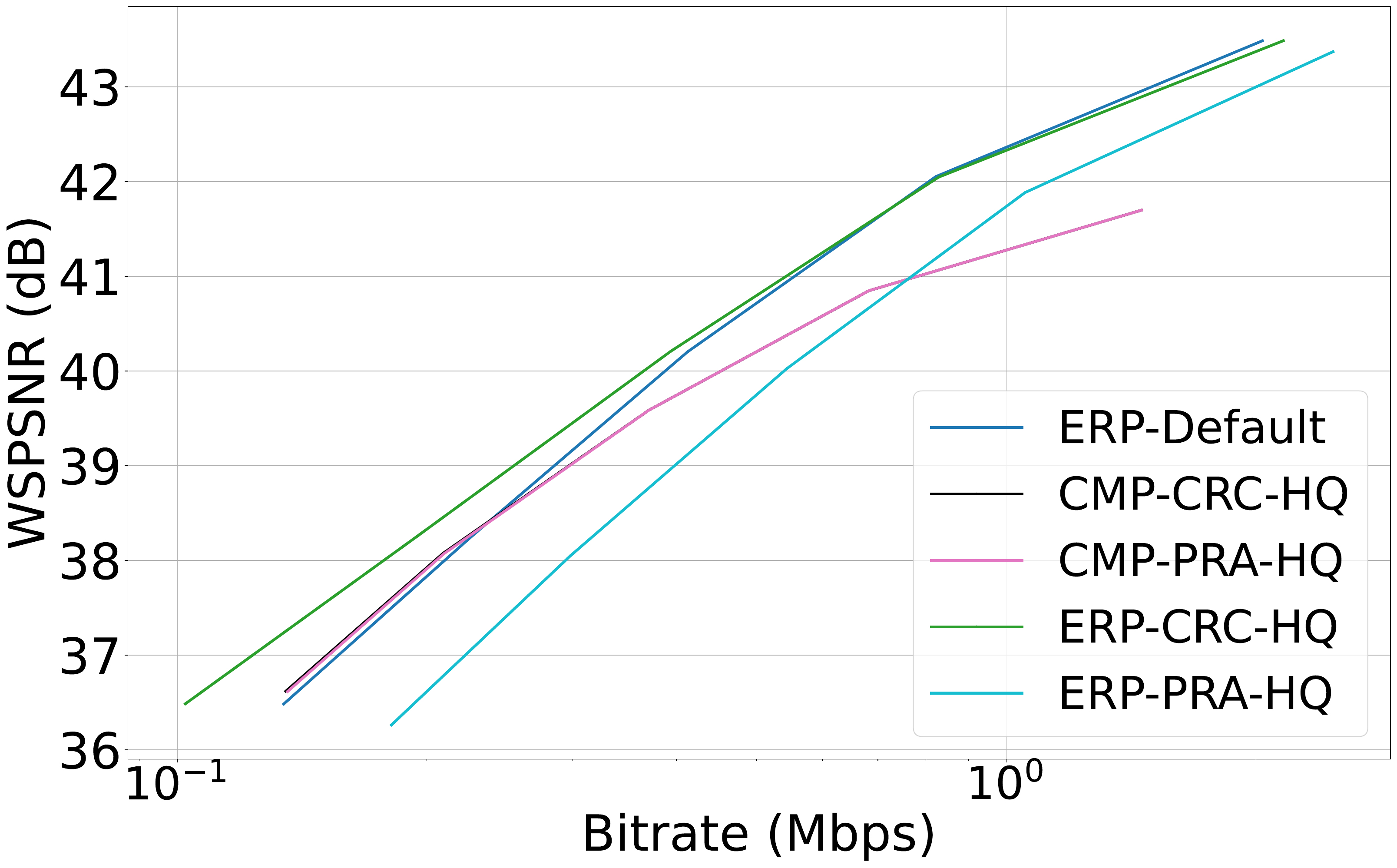}
\includegraphics[width=0.45\columnwidth]{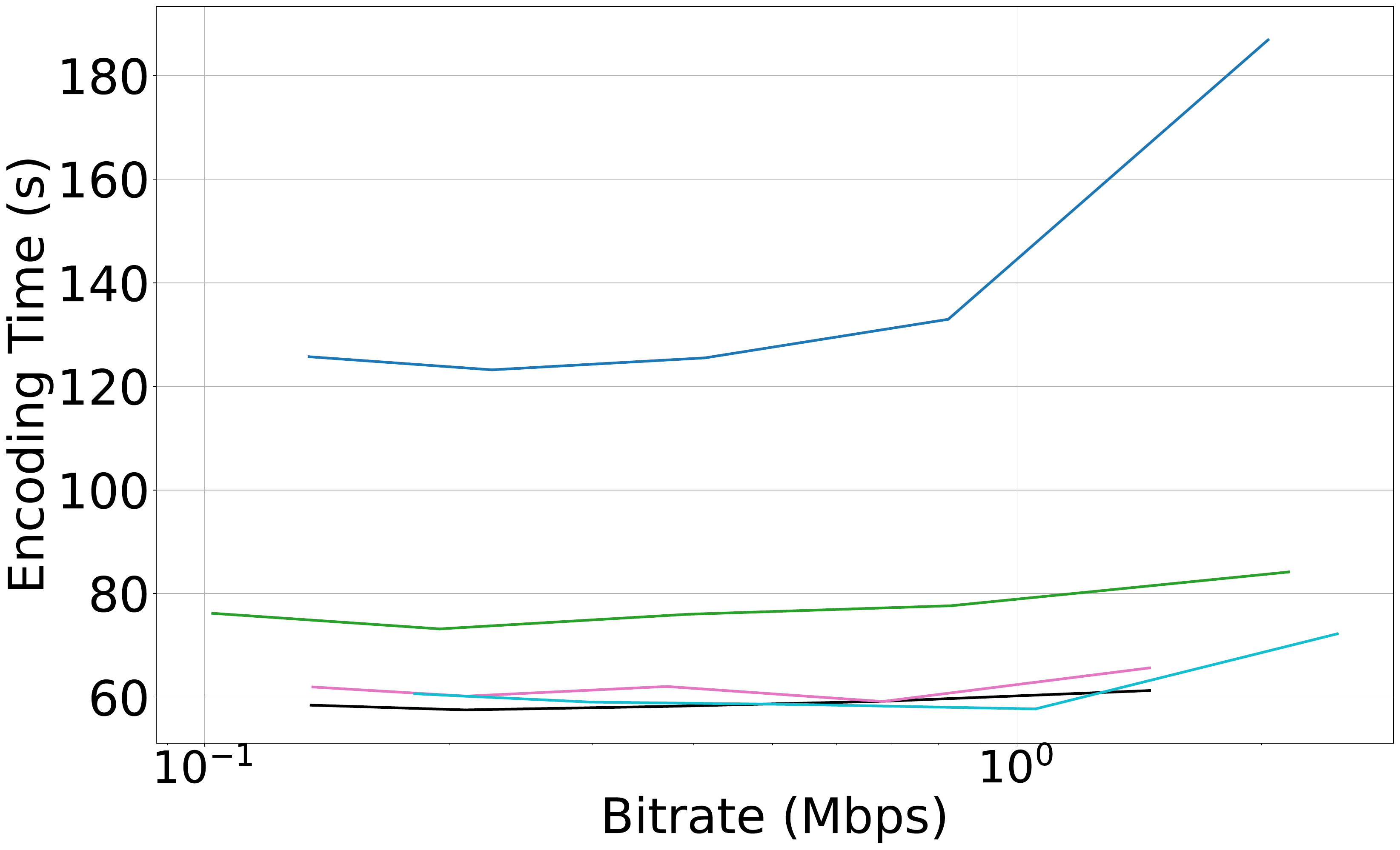}
\caption{4K}
\end{subfigure}

\begin{subfigure}{0.49\textwidth}
\centering
\includegraphics[width=0.45\columnwidth]{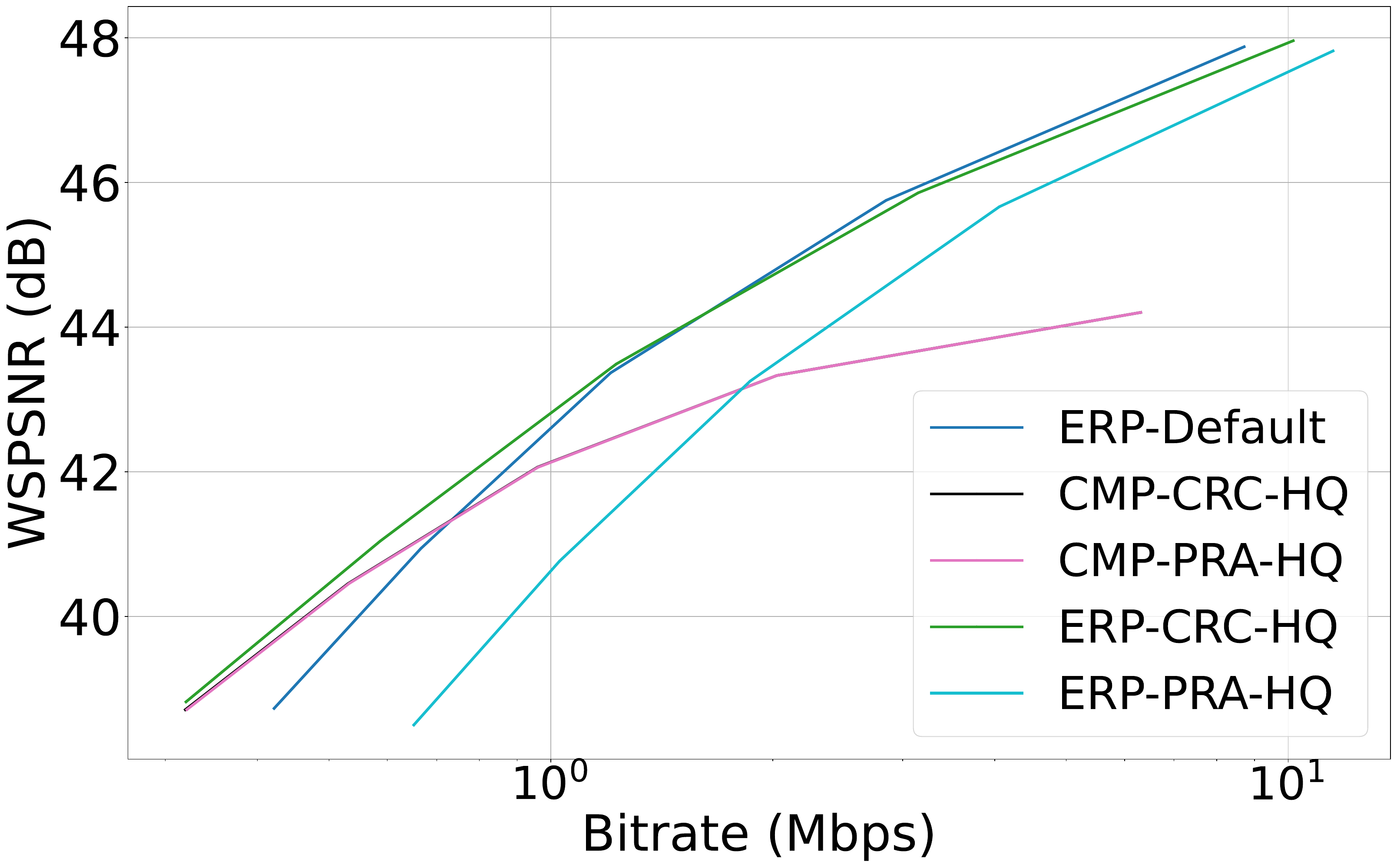}
\includegraphics[width=0.45\columnwidth]{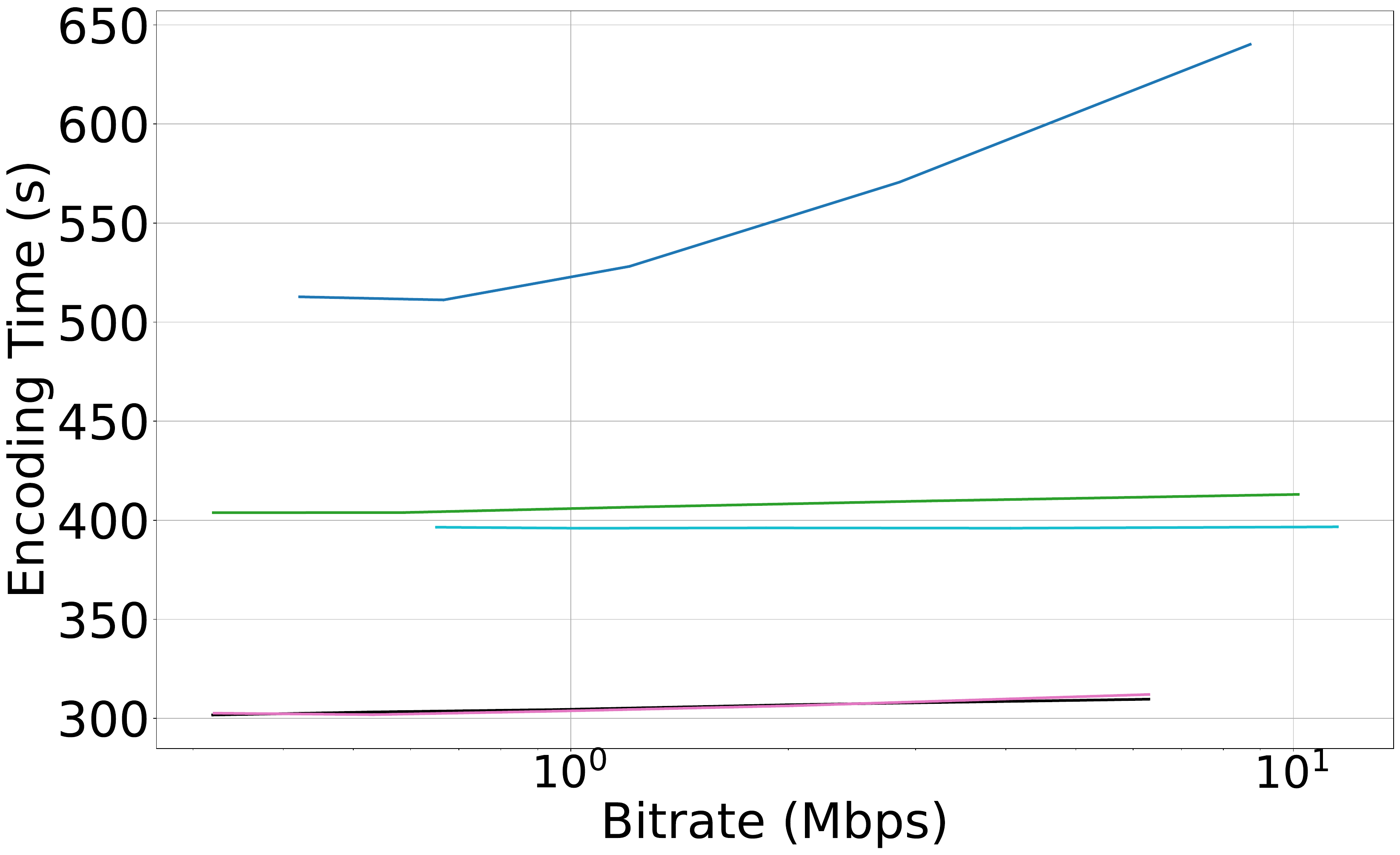}
\caption{8K}
\end{subfigure}
\caption{Rate--WSPSNR and rate--encoding time curves for \emph{Academic Building} at HD, 4K, and 8K.}
\vspace{-0.7em}
\label{fig:encT_res}
\end{figure}

\paragraph*{Encoding-Speed Gains} 
Figure~\ref{fig:encT_res} compares WSPSNR and encoding time across HD, 4K, and 8K for the \emph{Academic Building} sequence. The left plots show rate--WSPSNR curves, while the right plots show encoding time versus bitrate. ERP-Default yields the highest encoding times at all resolutions, reflecting the cost of exhaustive full-search encoding. Both \CRC and \PRA provide substantial speedups, with \PRA showing the most aggressive reductions. CMP-based variants (\mbox{CMP-\CRC} and \mbox{CMP-\PRA}) amplify these gains through face-wise parallelism, offering the greatest benefits at 8K where ERP-Default exceeds \SI{600}{\second} per representation. In terms of quality, \CRC follows ERP-Default most closely, whereas \PRA deviates slightly—capturing the fundamental trade-off between speed and RD fidelity. At 4K, \PRA provides the largest improvements, and CMP variants deliver the strongest overall throughput, highlighting the advantage of projection-aware reuse.

\begin{table*}[t]
\caption{Comparison of rate–distortion and encoding time savings for proposed methods relative to the ERP-Default encoding.}
\centering
\resizebox{0.7095\textwidth}{!}{
\begin{tabular}{l|c|c||c|c|c|c|c|c}
\specialrule{.12em}{.05em}{.05em}
\specialrule{.12em}{.05em}{.05em}
Method & Ref & Res & BD-PSNR & BD-WSPSNR & BDET & BDET & $\Delta T_S$ & $\Delta T_P$ \\
 & &  & [dB] & [dB] & (PSNR) [\%] & (WS-PSNR) [\%] & [\%] & [\%] \\
\specialrule{.12em}{.05em}{.05em}
\specialrule{.12em}{.05em}{.05em}

\multirow{12}{*}{ERP-\CRC}
& \multirow{4}{*}{LQ} & HD
& $-0.06 \pm 0.03$ & $-0.06 \pm 0.03$
& $-47.09 \pm 4.89$ & $-47.07 \pm 4.89$
& $-43.63 \pm 5.36$ & $-24.20 \pm 5.76$ \\
& & 4K
& $0.09 \pm 0.06$ & $0.08 \pm 0.06$
& $-50.43 \pm 1.68$ & $-50.43 \pm 1.66$
& $-51.57 \pm 1.20$ & $-56.64 \pm 4.24$ \\
& & 8K
& $0.02 \pm 0.13$ & $-0.02 \pm 0.14$
& $-29.63 \pm 5.22$ & $-29.67 \pm 5.20$
& $-31.04 \pm 5.45$ & $-40.84 \pm 5.87$ \\
& & Avg
& \cellcolor{gray!15}\textbf{0.02 $\pm$ 0.07} 
& \cellcolor{gray!15}\textbf{0.00 $\pm$ 0.08}
& \cellcolor{gray!15}\textbf{-42.38 $\pm$ 3.93} 
& \cellcolor{gray!15}\textbf{-42.39 $\pm$ 3.92}
& \cellcolor{gray!15}\textbf{-42.08 $\pm$ 4.01} 
& \cellcolor{gray!15}\textbf{-40.56 $\pm$ 5.29} \\
\cline{2-9}

& \multirow{4}{*}{MQ} & HD
& $-0.01 \pm 0.02$ & $-0.01 \pm 0.02$
& $-41.47 \pm 4.00$ & $-41.44 \pm 3.99$
& $-42.86 \pm 4.30$ & $-23.56 \pm 3.80$ \\
& & 4K
& $0.12 \pm 0.06$ & $0.10 \pm 0.06$
& $-48.29 \pm 2.09$ & $-48.28 \pm 2.10$
& $-49.45 \pm 2.64$ & $-54.32 \pm 4.81$ \\
& & 8K
& $0.03 \pm 0.12$ & $0.00 \pm 0.14$
& $-29.43 \pm 5.44$ & $-29.47 \pm 5.42$
& $-30.86 \pm 5.61$ & $-40.85 \pm 5.88$ \\
& & Avg
& \cellcolor{gray!15}\textbf{0.05 $\pm$ 0.07} 
& \cellcolor{gray!15}\textbf{0.03 $\pm$ 0.07}
& \cellcolor{gray!15}\textbf{-39.73 $\pm$ 3.84} 
& \cellcolor{gray!15}\textbf{-39.73 $\pm$ 3.84}
& \cellcolor{gray!15}\textbf{-41.06 $\pm$ 4.18} 
& \cellcolor{gray!15}\textbf{-39.58 $\pm$ 4.83} \\
\cline{2-9}

& \multirow{4}{*}{HQ} & HD
& $-0.05 \pm 0.05$ & $-0.05 \pm 0.06$
& $-40.30 \pm 5.92$ & $-40.28 \pm 5.92$
& $-33.33 \pm 7.06$ & $0.18 \pm 11.62$ \\
& & 4K
& $0.13 \pm 0.04$ & $0.12 \pm 0.03$
& $-44.29 \pm 2.69$ & $-44.28 \pm 2.69$
& $-45.57 \pm 2.97$ & $-50.41 \pm 7.06$ \\
& & 8K
& $0.06 \pm 0.09$ & $0.03 \pm 0.10$
& $-27.76 \pm 3.93$ & $-27.80 \pm 3.90$
& $-29.22 \pm 4.46$ & $-38.55 \pm 4.85$ \\
& & Avg
& \cellcolor{gray!15}\textbf{0.05 $\pm$ 0.06} 
& \cellcolor{gray!15}\textbf{0.03 $\pm$ 0.06}
& \cellcolor{gray!15}\textbf{-37.45 $\pm$ 4.18} 
& \cellcolor{gray!15}\textbf{-37.45 $\pm$ 4.17}
& \cellcolor{gray!15}\textbf{-36.04 $\pm$ 4.83} 
& \cellcolor{gray!15}\textbf{-29.59 $\pm$ 7.84} \\
\hline

\multirow{12}{*}{ERP-\PRA}
& \multirow{4}{*}{LQ} & HD
& $-0.19 \pm 0.14$ & $-0.18 \pm 0.11$
& $-54.28 \pm 4.69$ & $-54.25 \pm 4.73$
& $-50.62 \pm 4.86$ & $-25.00 \pm 5.22$ \\
& & 4K
& $-0.26 \pm 0.15$ & $-0.26 \pm 0.14$
& $-55.99 \pm 2.00$ & $-56.10 \pm 1.97$
& $-59.06 \pm 0.79$ & $-63.61 \pm 3.59$ \\
& & 8K
& $-0.44 \pm 0.19$ & $-0.41 \pm 0.23$
& $-27.02 \pm 3.63$ & $-27.21 \pm 3.68$
& $-32.35 \pm 5.18$ & $-42.62 \pm 5.57$ \\
& & Avg
& \cellcolor{gray!15}\textbf{-0.30 $\pm$ 0.16} 
& \cellcolor{gray!15}\textbf{-0.28 $\pm$ 0.16}
& \cellcolor{gray!15}\textbf{-45.77 $\pm$ 3.44} 
& \cellcolor{gray!15}\textbf{-45.86 $\pm$ 3.46}
& \cellcolor{gray!15}\textbf{-47.34 $\pm$ 3.61} 
& \cellcolor{gray!15}\textbf{-43.74 $\pm$ 4.79} \\
\cline{2-9}

& \multirow{4}{*}{MQ} & HD
& $-0.16 \pm 0.03$ & $-0.15 \pm 0.03$
& $-47.29 \pm 5.52$ & $-47.25 \pm 5.49$
& $-48.62 \pm 5.16$ & $-19.46 \pm 6.89$ \\
& & 4K
& $-0.21 \pm 0.08$ & $-0.22 \pm 0.07$
& $-55.97 \pm 1.42$ & $-56.03 \pm 1.38$
& $-58.78 \pm 1.15$ & $-64.31 \pm 3.36$ \\
& & 8K
& $-0.45 \pm 0.20$ & $-0.48 \pm 0.23$
& $-29.49 \pm 4.35$ & $-29.67 \pm 4.36$
& $-32.39 \pm 5.17$ & $-42.71 \pm 5.50$ \\
& & Avg
& \cellcolor{gray!15}\textbf{-0.27 $\pm$ 0.10} 
& \cellcolor{gray!15}\textbf{-0.29 $\pm$ 0.11}
& \cellcolor{gray!15}\textbf{-44.25 $\pm$ 3.76} 
& \cellcolor{gray!15}\textbf{-44.31 $\pm$ 3.74}
& \cellcolor{gray!15}\textbf{-46.59 $\pm$ 3.83} 
& \cellcolor{gray!15}\textbf{-42.16 $\pm$ 5.25} \\
\cline{2-9}

& \multirow{4}{*}{HQ} & HD
& $-0.54 \pm 0.08$ & $-0.55 \pm 0.10$
& $-54.39 \pm 4.14$ & $-54.34 \pm 4.14$
& $-46.86 \pm 5.66$ & $-11.03 \pm 11.02$ \\
& & 4K
& $-0.76 \pm 0.14$ & $-0.79 \pm 0.14$
& $-55.91 \pm 1.65$ & $-55.92 \pm 1.65$
& $-56.66 \pm 1.67$ & $-58.37 \pm 5.78$ \\
& & 8K
& $-1.18 \pm 0.37$ & $-1.26 \pm 0.42$
& $-30.49 \pm 4.51$ & $-30.56 \pm 4.48$
& $-31.89 \pm 4.69$ & $-41.13 \pm 5.13$ \\
& & Avg
& \cellcolor{gray!15}\textbf{-0.83 $\pm$ 0.20} 
& \cellcolor{gray!15}\textbf{-0.87 $\pm$ 0.22}
& \cellcolor{gray!15}\textbf{-46.93 $\pm$ 3.43} 
& \cellcolor{gray!15}\textbf{-46.94 $\pm$ 3.42}
& \cellcolor{gray!15}\textbf{-45.14 $\pm$ 4.01} 
& \cellcolor{gray!15}\textbf{-36.84 $\pm$ 7.31} \\
\hline

\multirow{4}{*}{CMP-\CRC} & \multirow{4}{*}{HQ}
& HD
& $-0.60 \pm 0.31$ & $-0.29 \pm 0.43$
& $-49.56 \pm 6.03$ & $-50.24 \pm 5.72$
& $-46.33 \pm 6.54$ & $-27.66 \pm 7.25$ \\
& & 4K
& $-0.92 \pm 0.08$ & $-0.56 \pm 0.19$
& $-54.92 \pm 1.84$ & $-55.11 \pm 1.98$
& $-58.01 \pm 1.81$ & $-62.96 \pm 4.87$ \\
& & 8K
& $-1.70 \pm 0.88$ & $-1.44 \pm 0.77$
& $-43.77 \pm 4.23$ & $-43.99 \pm 4.17$
& $-48.12 \pm 4.23$ & $-55.40 \pm 4.51$ \\
& & Avg
& \cellcolor{gray!15}\textbf{-1.07 $\pm$ 0.42} 
& \cellcolor{gray!15}\textbf{-0.76 $\pm$ 0.46}
& \cellcolor{gray!15}\textbf{-49.42 $\pm$ 4.04} 
& \cellcolor{gray!15}\textbf{-49.78 $\pm$ 3.96}
& \cellcolor{gray!15}\textbf{-50.82 $\pm$ 4.19} 
& \cellcolor{gray!15}\textbf{-48.67 $\pm$ 5.54} \\
\hline

\multirow{4}{*}{CMP-\PRA} & \multirow{4}{*}{HQ}
& HD
& $-0.60 \pm 0.31$ & $-0.29 \pm 0.43$
& $-50.41 \pm 4.24$ & $-51.05 \pm 4.00$
& $-47.59 \pm 4.46$ & $-30.76 \pm 4.22$ \\
& & 4K
& $-0.94 \pm 0.07$ & $-0.57 \pm 0.18$
& $-55.24 \pm 2.79$ & $-55.40 \pm 2.90$
& $-58.28 \pm 1.90$ & $-63.46 \pm 4.20$ \\
& & 8K
& $-1.70 \pm 0.89$ & $-1.45 \pm 0.77$
& $-43.98 \pm 4.14$ & $-44.19 \pm 4.09$
& $-48.27 \pm 4.17$ & $-55.43 \pm 4.54$ \\
& & Avg
& \cellcolor{gray!15}\textbf{-1.08 $\pm$ 0.42} 
& \cellcolor{gray!15}\textbf{-0.77 $\pm$ 0.46}
& \cellcolor{gray!15}\textbf{-49.88 $\pm$ 3.73} 
& \cellcolor{gray!15}\textbf{-50.22 $\pm$ 3.66}
& \cellcolor{gray!15}\textbf{-51.38 $\pm$ 3.51} 
& \cellcolor{gray!15}\textbf{-49.88 $\pm$ 4.32} \\
\specialrule{.12em}{.05em}{.05em}
\specialrule{.12em}{.05em}{.05em}
\end{tabular}
}
\label{tab:avg_results}
\end{table*}

Table~\ref{tab:avg_results} summarizes performance relative to ERP-Default. For ERP content, \CRC reduces serial time by \SI{33}{\percent}--\SI{52}{\percent} and parallel time by \SI{24}{\percent}--\SI{57}{\percent}. CMP-\CRC further improves these numbers, reaching \SI{50.82}{\percent} (serial) and \SI{48.67}{\percent} (parallel). \PRA achieves the largest raw reductions—\SI{45}{\percent}--\SI{47}{\percent} on average and up to \SI{59}{\percent} at 4K—while CMP-\PRA delivers the highest overall throughput, with average gains of \SI{51.38}{\percent} (serial) and \SI{49.88}{\percent} (parallel). These trends demonstrate that bottom-up reuse remains highly effective even under the more complex statistics of spherical video.  Such savings are particularly valuable in production pipelines where dozens of representations must be prepared per sequence.

\emph{Rate-Distortion Impact: } 
\CRC offers the most stable RD behavior, with ERP BD-PSNR/BD-WSPSNR differences close to zero. CMP-\CRC exhibits larger losses (up to \SI{-1.07}{\decibel} BD-PSNR), and \PRA yields more pronounced deviations due to its more aggressive reuse granularity.
However, WSPSNR accounts for spherical sampling nonuniformity, and hence, it more closely reflects perceptual differences than planar PSNR in 360° content. Thus, the largest CMP–CRC drop at 8K appears primarily in planar PSNR; corresponding WSPSNR differences are considerably smaller and remain visually acceptable for typical VR viewports, where spherical distortion patterns and user-centric masking reduce perceptibility. 

\emph{Complexity and Resource Considerations: } 
For $N$ resolutions, \CRC stores analysis $(N\!-\!1)$ times, whereas \PRA stores $N$ anchors. Thus, \CRC is somewhat more memory-efficient, though PRA’s footprint remains manageable for modern servers. The comparison underscores a practical trade-off: CRC favors conservative reuse and compact resource use, while PRA sacrifices small amounts of RD precision for maximum throughput.

\emph{CMP-Based Variants and Practical Implications: } 
CMP-\CRC and CMP-\PRA follow the ERP trends but benefit from independent face-wise processing, enabling the highest overall speedups (up to $4.2\times$). The additional parallelism, combined with more uniform face statistics, makes CMP-based pipelines particularly attractive for time-sensitive applications such as live VR streaming or rapid multi-representation preparation in large content libraries. These results, to our knowledge, constitute the first systematic comparison of analysis‑reuse behavior under ERP and CMP projections within OMAF‑compliant multirate encoding pipelines.

\section{Conclusions and Future Work}
\label{sec:conc}
This paper presented a fast multirate encoding framework for cubemap $360^{\circ}$ video in OMAF-compliant streaming workflows. Building on x265’s analysis–reuse capabilities, we introduced two cross-resolution strategies—Cascaded Resolution Chain (\CRC) and Per-Resolution Anchoring (\PRA)—along with their cubemap-based counterparts, CMP-\CRC and CMP-\PRA. Using the SJTU 8K $360^{\circ}$ dataset, our experiments demonstrated that all variants achieve substantial encoding-time reductions compared to exhaustive full-search encoding, with \mbox{BDET} of up to \SI{-50}{\percent} and wall-clock speedups reaching $4.2\times$ in the most favorable CMP-\CRC configurations. Our results highlight a clear trade-off: \CRC offers better RD fidelity and lower memory footprint (analysis stored $N-1$ times), while \PRA maximizes encoding speed at the cost of modest RD degradation and slightly higher memory overhead ($N$ anchors). Our CMP-based designs further exploit face-wise independence to enable parallel encoding, delivering up to 4.2× speedups and making the framework suitable for near-real-time scenarios such as live VR streaming. Importantly, these enhancements integrate seamlessly with OMAF’s region-based architecture and DASH delivery without requiring client-side modifications.  

Future work includes extending the proposed framework to VVC-based pipelines, exploring content-aware anchor selection and reuse strength (\eg via machine learning or online complexity classification), and integrating viewport prediction or QoE-driven bitrate ladder optimization. Another promising direction is to combine the proposed encoder-side acceleration with decoder-side enhancements, such as super-resolution or view-adaptive upsampling, to enable next-generation immersive streaming services.
\balance
\newpage
\bibliographystyle{IEEEtran}
\bibliography{references.bib,360degree}
\balance
\end{document}